\documentclass[a4paper,fleqn]{cas-sc}
\usepackage[numbers]{natbib}
\def\tsc#1{\csdef{#1}{\textsc{\lowercase{#1}}\xspace}}
\tsc{WGM}
\tsc{QE}
\tsc{EP}
\tsc{PMS}
\tsc{BEC}
\tsc{DE}
\usepackage{stackrel}

\newcommand{\sugb}[1]{\textcolor{blue}{#1}}

\begin{document}
\let\WriteBookmarks\relax
\def\floatpagepagefraction{1}
\def\textpagefraction{.001}
\shorttitle{Effect of the Förster Interaction and the Pulsed Pumping on the Quantum Correlations \dots}
\shortauthors{Madrid et~al.}
%\begin{frontmatter}

\title [mode = title]{Effect of the Förster Interaction and the Pulsed Pumping on the Quantum Correlations of a Two Quantum Dot-Microcavity System in the Strong Coupling Regime}                      

\author[1]{D. Madrid-Úsuga}[
orcid=https://orcid.org/0000-0002-4916-0211]
\cormark[1]%Corresponding author
%\fnmark[]
\ead{duvalier.madrid@unisucrevirtual.edu.co}
%\ead[URL]{}
\credit{Conceptualization, Numerical simulations and writing of the manuscript}
\address[1]{Departamento de F\'{i}sica, Universidad de Sucre, A.A. 406, Sincelejo, Colombia.}

\author[2]{ A. A. Portacio}%[
\credit{numerical simulations and writing of the manuscript}
\address[1]{Universidad de los Llanos, Facultad de Ciencias B\'{a}sicas e Ingenier\'{i}a, Villavicencio, Colombia.} 

\author[3]{ D. A. Rasero}%[
\credit{Numerical simulations and writing of the manuscript}
\address[3]{Departamento de Ciencias Naturales, Grupo de F\'{i}sica Aplicada FIASUR, Universidad Surcolombiana, A.A 385, Neiva, Colombia.}

\cortext[cor2]{Principal corresponding author}

\begin{abstract}
The quantum correlations of a system of two quantum dots with Föster interaction ($\Gamma$) in a microcavity with strongly coupled dissipation and a single mode of the electromagnetic field and driven by a laser pulse were studied theoretically, using the formalism of the master equation in Lindbland form. The energy eigenvalues of the system were studied as a function of detuning for the first and second excitation varieties. Concurrence ($\mathcal{CC}$), formation entanglement ($\mathcal{E}o\mathcal{F}$), mutual information~($\mathcal{I}$) and quantum discord ($\mathcal{Q}$) are studied as a function of time considering different values of Föster coupling, varying the pump times of the simulated laser pulse and pulse intensity. We found a discrepancy between $\mathcal{E}o\mathcal{F}$ and $\mathcal{CC}$ as entanglement quantifiers, noting that concurrence reaches much higher values than $\mathcal{E}o\mathcal{F}$; so concurrence can indicate results that are well above the $\mathcal{E}o\mathcal{F}$. The presence of the Föster interaction favors that the quantum discord is the dominant correlation in the system, which indicates that the system maintains quantum correlations even when the entanglement of the system has disappeared, but that it is affected by the increase in the laser pump time.\\
\end{abstract}

\begin{graphicalabstract}
\includegraphics[keepaspectratio, width = 160mm]{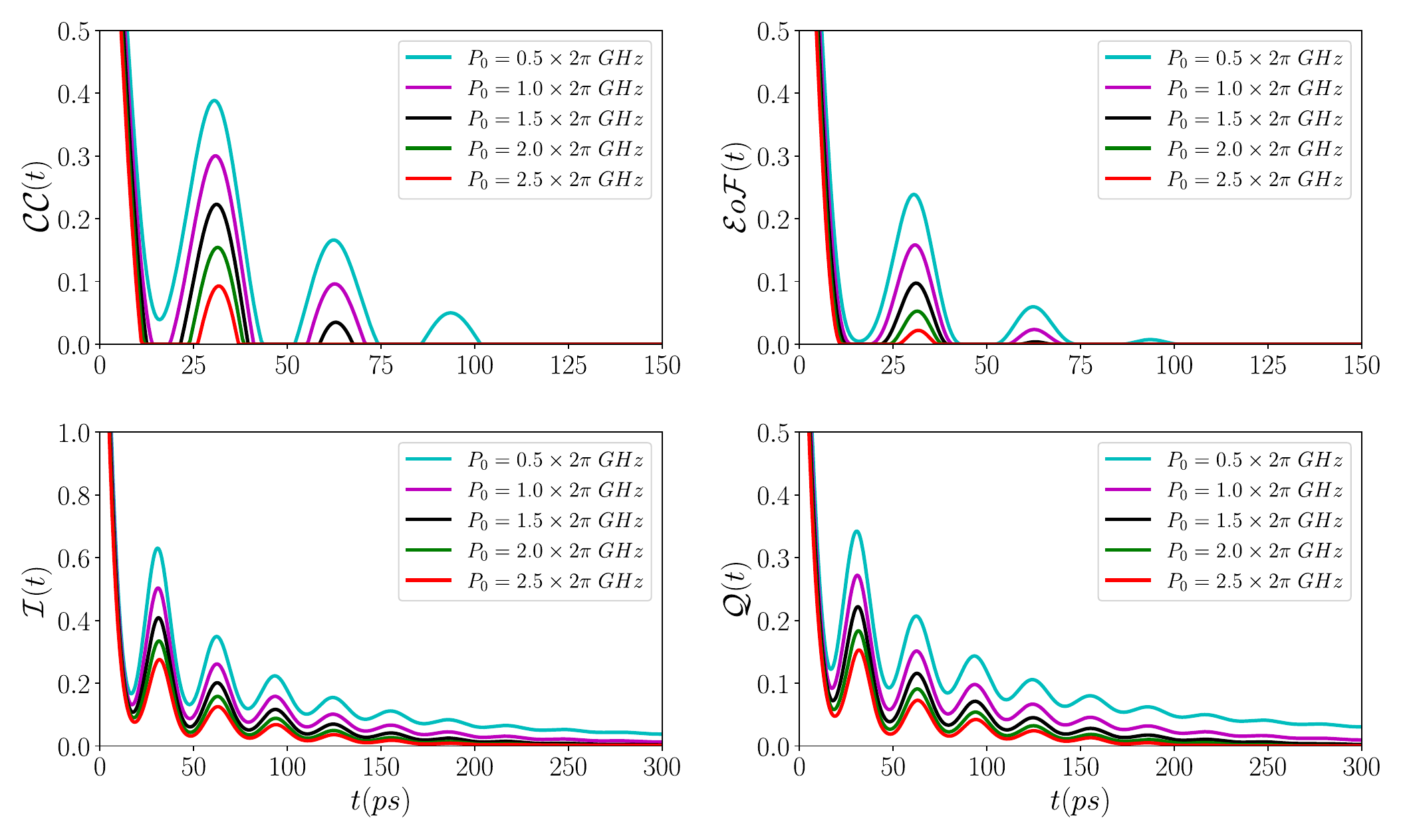}
\end{graphicalabstract}

\begin{highlights}
\item Quantum correlations were calculated as a function of time for different values of Förster interaction, electromagnetic field intensity and laser pulse time.
\item A discrepancy between $\mathcal{E}o\mathcal{F}$ and concurrence as entanglement quantifiers is shown, observing that concurrence reaches much higher values than $\mathcal{E}o\mathcal{F}$.
\end{highlights}


\maketitle
\section{Introduction}

The quantum properties of physical systems have been studied for many years as crucial resources for processing tasks, quantum control and non-locality of quantum information protocols, and correlations between quantum objects as fundamental characteristics of systems~\sugb{\cite{knill2010,ladd2010q,bennett2000}}. The coupling between quantum dot (QD), for example, is a phenomenon that has acquired fundamental importance in quantum logic devices and quantum computing, where QDs interact through the well-known Förster interaction, which considers a resonant energy transfer. This interaction has been experimentally evidenced in the dynamics of energy transfer in ensembles of QDs~\sugb{\cite{crooker2002}} and electron transfer of quantum dots of CdSe~\sugb{\cite{kagan1996}}. In addition, it has allowed us to obtain and learn relevant information on electro-optical and entanglement properties~\sugb{\cite{nazir2005, wang2021, guhne2009, friis2019}}. 

The study of such properties in open quantum systems contributes to the construction of a clearer and more precise horizon on quantum information and its field of applications~\sugb{\cite{prokopenko2014, Reina2014}}, particularly because properties such as decoherence appear as a ubiquitous physical process that prevents the realization of unitary quantum dynamics, eliminates quantum coherence and multipartite correlation effects, and has long been recognized as a mechanism responsible for the emergence of event classicism in a purely quantum realm~\sugb {\cite{zurek2003, zurek2007}}. To understand these properties, the use of the QD-QD interaction parameter is interesting, since it allows generating a quantum information processing scheme, through the use of the dynamics of quantum correlations, in which quantum entanglement~\sugb{\cite{wootters1998, wootters1998A, horodecki2009, chen2021}}, has become a key quantifier of the correlations present in various types of quantum systems~\sugb{\cite{reina2014, susa2014}}. However, all the quantum correlations present in the system cannot be quantified simply by entanglement, as has been studied in several works~\sugb{\cite{shajilal2023, kenfack2017,essakhi2022}}. Therefore, Oliver and Zurek introduce the concept of quantum discord ($\mathcal{Q}$) as a quantifier of all quantum correlations~\sugb{\cite{liu2022, modi2012}}; since it can exist even for some separable mixed states.

The study of such properties in open quantum systems contributes to the construction of a clearer and more precise horizon on quantum information and its field of applications~\sugb{\cite{prokopenko2014, Reina2014}}, particularly because properties such as decoherence appear as a ubiquitous physical process that prevents the realization of unitary quantum dynamics, eliminates quantum coherence and multipartite correlation effects, and has long been recognized as a mechanism responsible for the emergence of event classicism in a purely quantum realm~\sugb {\cite{zurek2003, zurek2007}}. To understand these properties, the use of the QD-QD interaction parameter is interesting, since it allows generating a quantum information processing scheme, through the use of the dynamics of quantum correlations, in which quantum entanglement~\sugb{\cite{wootters1998, wootters1998A, horodecki2009, chen2021}}, has become a key quantifier of the correlations present in various types of quantum systems~\sugb{\cite{reina2014, susa2014}}. However, all the quantum correlations present in the system cannot be quantified simply by entanglement, as has been studied in several works~\sugb{\cite{shajilal2023, kenfack2017,essakhi2022}}. Therefore, Oliver and Zurek introduce the concept of quantum discord ($\mathcal{Q}$) as a quantifier of all quantum correlations~\sugb{\cite{liu2022, modi2012}}; since it can exist even for some separable mixed states.

In this work, we consider a 2QD-Cavity system with dipole-dipole interaction, using the formalism of the master equation in the Lindblad form. The system is modeled with the Tavis-Cummings Hamiltonian, which describes the interaction of the QD with the cavity mode. The Lindbladian includes terms of spontaneous emission, cavity decay, incoherent pumping of the cavity and a pulsed excitation. The pumping scheme that is used in these works typically consider a continuous-wave pumping. However, in order to achieve the high densities required for the quantum strong coupling regimen (QSC), experiments typically use a pulsed excitation due to the power constraints of the typical lasers used. Such pulsed excitations give a time-dependent, non-equilibrium element to the problem. The dynamics of the system is determined not only by the Hamiltonian of system, but due to the open-dissipative nature of the matter and bosonic fields, which have a finite decay time. Therefore, in this work we consider a regime where the duration of the pulse is shorter than the lifetime of the particles in the system. Due to the pulsed excitation, the system is, in general, always in a non-equilibrium state. this allows the possibility of directly studying the system during the decomposition process through the observation of quatic correlations.

This article is organized as follows: In section \sugb{\ref{system}}, a description of the system considered for the study is made, in section \sugb{\ref{Master_equation}}, the master equation that describes the unitary and non-unitary evolution is presented, as well as the terms of dissipation and pumping involved in the process, Section \sugb{\ref{correlations}}, presents a brief description of the theoretical framework necessary to perform the calculations corresponding to quantum correlations. Section \sugb{\ref{discusion}} presents the results and a discussion of them. Finally, in section \sugb{\ref{Conclusions}}, we present our conclusions.

\section{The Considered System}\label{system}
\begin{figure}[t]
\centering
\includegraphics[scale=0.9]{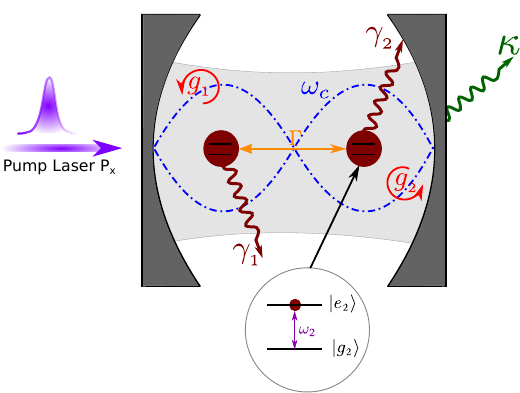}
\caption{Representation of the coupled 2-QD-microcavity system with a single mode of the quantized electromagnetic field forced by a laser pulse. The particles, oscillating between the ground state $\vert g \rangle_i$ and the excited state $\vert e \rangle_i$, are pumped into the cavity by an  pump laser $P_x$. The excited particles in the cavity disintegrate with losses due to leakage of photons through the cavity $\kappa$ or spontaneous emission $\gamma_i$ of the 2-QD's. The coupling force between the cavity photon and a two-level atom is given by $g_i$.}
\label{Fig_1}
\end{figure}
The proposed system consists of two quantum dots interacting with a confined electromagnetic single-mode of a microcavity. Which are coupled with each other through a dipole-dipole interaction, and with the field through a Jaynes-Cummings interaction. Each QD is considered as a two-level system, so that a QD may be in the ground state or an excited state. Fig.~\sugb{\ref{Fig_1}} shows the diagram of the physical system to be considered, with its respective mechanisms of loss and pumping of the QD's-nanocavity system. The two-level system can be, in principle, replaced by ions, atoms, quantum dots, superconducting qubits, etc. In this configuration, the Hamiltonian can be given by~\sugb{\cite{zhang2014,nicolosi2004}}
\begin{equation}
\mathcal{H}_{s}=\hbar\omega_ca^{\dagger}a + \sum_{i=1}^2\hbar\omega_i\sigma_+^i\sigma_-^i + \sum_{i=1}^2\hbar g_i(a^{\dagger}\sigma_-^i + a\sigma_+^i) +\hbar\Gamma(\sigma_+^1\sigma_-^2 + \sigma_-^1\sigma_+^2)
\label{Eq_1}
\end{equation}
where $\sigma_{+}^i=\vert e_i\rangle\langle g_i\vert$ and $\sigma_{-}^i=\vert g_i\rangle\langle e_i\vert$ are, respectively, the excitonic operators for each of the quantum dots, $\omega_i$ is the frequency of \textit{i-th} QD. Note that the symbols $\vert e \rangle$ and $\vert g \rangle$ refer to the excited and ground states for the QD’s. Here, $a^{\dagger}$ ($a$) is the bosonic operator of creation (destruction) that generates the photonic states in the cavity, which is characterized by the frequency $\omega_c$, the terms $g_i$ correspond to the light-matter interaction constants between the QD's and the cavity mode and~$\Gamma$ corresponds to the Förster exciton-exciton interaction. We assume that the transition frequencies of QD$_1$ and QD$_2$ are equal $\omega_1=\omega_2=\omega_c + \Delta$ (where $\Delta$ is the detuning of each exciton with respect to the field mode). And  both coupling coefficients between the QDs and the field are $g$ ($g_1=g_2=g$). The system is characterized by the decay of cavity energy and the spontaneous emission of quantum dots $\lbrace\kappa, \gamma\rbrace/2\pi = \lbrace 5.0; 0.025\rbrace$~GHz, corresponding to quality factor $Q =\omega_c/2\kappa\approx 17000$, and the individual emitter-cavity coupling rate $g/2\pi = 10.0$~GHz, that guarantee the strong coupling regime condition: $g > \kappa, \gamma$~\sugb{\cite{reithmaier2004, yoshie2004, majumdar2012}}.

The Hamiltonian Eq.~(\sugb{\ref{Eq_1}}) can be written on the basis of naked states $\lbrace\vert g\rangle, \vert e\rangle\rbrace\otimes\lbrace\vert g\rangle, \vert e\rangle\rbrace \otimes \lbrace \vert n \rangle\rbrace_{n=0}^{\infty}$, as a diagonal block matrix. In this basis $\vert e\rangle$ the excited  and ground $\vert g\rangle$ states of the ith exciton, whereas $\vert n\rangle$ indicates a Fock state with $n$ photons. The matrix term corresponding to the \textit{n-th} block, that is, excitation manifolds $\Lambda_n$ is:

\begin{equation}
\mathcal{H}_s^{(n)}  =
\begin{bmatrix}
\hbar\omega_cn     & \hbar g_1\sqrt{n}                &     \hbar g_2\sqrt{n}            &           0                \\
\hbar g_1\sqrt{n}  & \hbar\omega_c(n-1)+\hbar\omega_1 &     \hbar\Gamma                  & \hbar g_2\sqrt{n-1}        \\
\hbar g_2\sqrt{n}  & \hbar\Gamma                      & \hbar\omega_c(n-1)+\hbar\omega_2 & \hbar g_1\sqrt{n-1}        \\
0                  & \hbar g_2\sqrt{n-1}              & \hbar g_1\sqrt{n-1}              & \hbar\omega_c(n-2)+\hbar\omega_1+\hbar\omega_2 \\
\end{bmatrix}
\end{equation}
\label{Eq_2}

with set of states $\Lambda_n=\lbrace\vert g,g,n\rangle,\vert g,e,n-1\rangle,\vert e,g,n-1 \rangle,\vert e,e,n-2\rangle\rbrace$.

\section{Master Equation of System}\label{Master_equation}

The Hamiltonian dynamics explained above ignores any effect of the loss. One way to incorporate the loss effect is to use the quantum optical master equation. For the system considered, there are two main sources of loss: dissipation of the cavity field to the environment with a loss rate $\kappa$ and a spontaneous emission rate $\gamma_i$ of the 2QD's. The dynamics of the QD cavity system, in the presence of losses, is determined by the master equation~\sugb{\cite{breuer2002,zhang2016}}:

\begin{align}\notag
\frac{d}{dt}\rho_{s} &= -\frac{i}{\hbar}[\mathcal{H}s,\rho_s]+\frac{\kappa}{2}(2a\rho a^{\dagger}-a^{\dagger}a\rho-\rho a^{\dagger}a) + \frac{\gamma}{2}\sum_{i=1}^2(2\sigma_i\rho\sigma_i^{\dagger}-\sigma_i^{\dagger}\sigma_i\rho-\rho\sigma_i^{\dagger}\sigma_i) \\
&+\frac{P_c}{2}(2a^{\dagger}\rho a-a a^{\dagger}\rho-\rho a a^{\dagger}) + \frac{P_x(t)}{2}\sum_{i=1}^2(2\sigma^{\dagger}_i\rho\sigma_i-\sigma_i\sigma^{\dagger}_i\rho-\rho\sigma_i\sigma^{\dagger}_i) 
\label{Eq_3}
\end{align}

The first describes the coherent evolution of the 2QD's-microcavity system, the second and third terms describe the decay of the cavity and the processes of spontaneous emission of the \textit{i-th} QD, respectively, and the fourth term denotes the pumping of photons and exciton. In this equation, $\kappa$ is the rate at which photons escape through the mirrors of the cavity, $\gamma_i$ is the rate of decay due to spontaneous emission from the QD's~\sugb{\cite{kopylov2015}}, the $P_{c}$ is the incoherent pumping of the cavity mode and $P_x(t)$ is the rate in wich excitons are created by the continuous pump laser to the higher order cavity mode, which obeys a Gaussian profile in the time domain with a pulse time~$\tau_p=\frac{FWHM}{2\sqrt{2\ln{2}}}$, with the form $P_{x}(t)=P_0 \cdot\exp\Big[\Big(\frac{(t-t_0)^2} {2\tau_p^2}\Big)\Big]$, where $P_0$ is proportional to the laser intensity, with $t_0$ the pulse central time. 

\section{Quantum Correlations}\label{correlations}

The definitions involved in the calculation of the correlations are as follows: Entanglement is a measure of the non-separability of the quantum state of a composite system and it is, in general, a difficult quantity to compute~\sugb{\cite{bennett2000, stephenson2020}} as a resource in quantum information and computation tasks, the entanglement expresses the maximum fractional number of Bell pair that it is possible to obtain from the quantum state to be used for quantum tasks. From the work of Wootters~\sugb{\cite{wootters1998}}, it is well known that the entanglement for a pair of qubits can be quantified by the concurrence. Therefore, the quantum entanglement is quantified by the $\mathcal{E}o\mathcal{F}$

\begin{equation}
\mathcal{E}o\mathcal{F}(\rho) = h\Bigg(\frac{1+\sqrt{1-\mathcal{CC}^2(\rho)}}{2} \Bigg),
\label{Eq_4}
\end{equation}
where $h(x)=-x\log_2 x-(1-x)\log_2(1-x)$ denotes the binary entropy function.
\medskip

Additionally, consider, in decreasing order, the eigenvalues $\lambda_i$ of the matrix $\sqrt{\rho_{_{AB}}\tilde{\rho}_{_{AB}}}$, where $\tilde{\rho}_{_{AB}}=(\sigma_y\otimes\sigma_y)\bar{\rho}_{_{AB}}(\sigma_y\otimes\sigma_y)$ and $\bar{\rho}_{_{AB}}$ is the element complex conjugate of $\rho_{_{AB}}$.
The concurrence $\mathcal{CC}$ can be defined as

\begin{equation}
\mathcal{CC}(\rho_{_{AB}}) = \max\lbrace{0,\lambda_1 -\lambda_2 - \lambda_3 - \lambda_4\rbrace},
\label{Eq_5}
\end{equation}

where the $\lambda_i$ are as introduced above or, equivalently (also in decreasing order), the square root of the eigenvalues of the non-Hermitian matrix $\rho_{_{AB}}\tilde{\rho}_{_{AB}}$~\sugb{\cite{wootters1998}}.

On the other hand, the quantum mutual information describes the whole content of correlations in a given quantum system~\sugb{\cite{groisman2005,modi2010}}. It has been shown that quantum correlations (entanglement included)~\sugb{\cite{ollivier2001}} and classical correlations~\sugb{\cite{henderson2001}}, in the sense of entropic measures, add up to give the quantum mutual information~\sugb{\cite{hamieh2004}}. Furthermore, this point has been recently emphasized, via the use of the relative entropy, within a unified framework that captures both quantum and classical correlations within the quantum mutual information~\sugb{\cite{modi2010}}. For a bipartite system, this can be written as

\begin{equation}
\mathcal{I}(\rho_{_{AB}})=\mathcal{S}(\rho_{_A})+\mathcal{S}(\rho_{_B})-\mathcal{S}(\rho_{_{AB}}),
\label{Eq_6}
\end{equation}

where $\mathcal{S}(\rho)=-Tr(\rho \log\rho)$ is the von Neumann entropy of density matrix $\rho$.
\medskip

A measure of classical correlations was introduced in~\sugb{\cite{henderson2001}} as the maximum extractable classical information from a subsystem A, when a set of positive operator-valued measures~\sugb{\cite{hamieh2004}} has been performed on the other subsystem~B:

\begin{equation}
\mathcal{C}(\rho_{_{AB}})=\stackbin[\lbrace\Pi_j^B\rbrace]{}{\sup}\Bigg[ S(\rho_A)-\sum_jp_jS(\rho_A^j)\Bigg],
\label{Eq_7}
\end{equation}

where $S(\rho_A^j)$ is the entropy associated with the density matrix of the subsystem A after the measure. Such correlations must be non-increasing, and invariant under local unitary operations, and $\mathcal{C}(\rho_{_{AB}})=0$ iff $\rho_{_{AB}}=\rho_{_A}\otimes \rho_{_B}$.
\medskip

Following the definition for $\mathcal{C}(\rho_{_{AB}})$, a simple way to define the total quantum correlations in a bipartite system is $\mathcal{Q}(\rho_{_{AB}})=\mathcal{I}(\rho_{_{AB}})-\mathcal{C}(\rho_{_{AB}})$. In terms of the von Neumann entropies, the quantum correlations, which coincide with the definition for the quantum discord given in \sugb{\cite{ollivier2001}}, read
\begin{equation}
\mathcal{Q}(\rho_{_{AB}})=S(\rho_{_B})-S(\rho_{_{AB}})+\stackbin[\lbrace\Pi_j^B\rbrace]{}{\inf}\sum_jp_jS(\rho_{A\vert \Pi_j^B}).
\label{Eq_8}
\end{equation}

For pure states, $\mathcal{Q}(\rho_{_{AB}})=S(\rho_{_B})$, and $\mathcal{Q}(\rho_{_{AB}})=0$ if the system is purely classically correlated.

\section{Results and Discussions} \label{discusion}

\begin{figure}[t]
\centering
\includegraphics[scale=0.5]{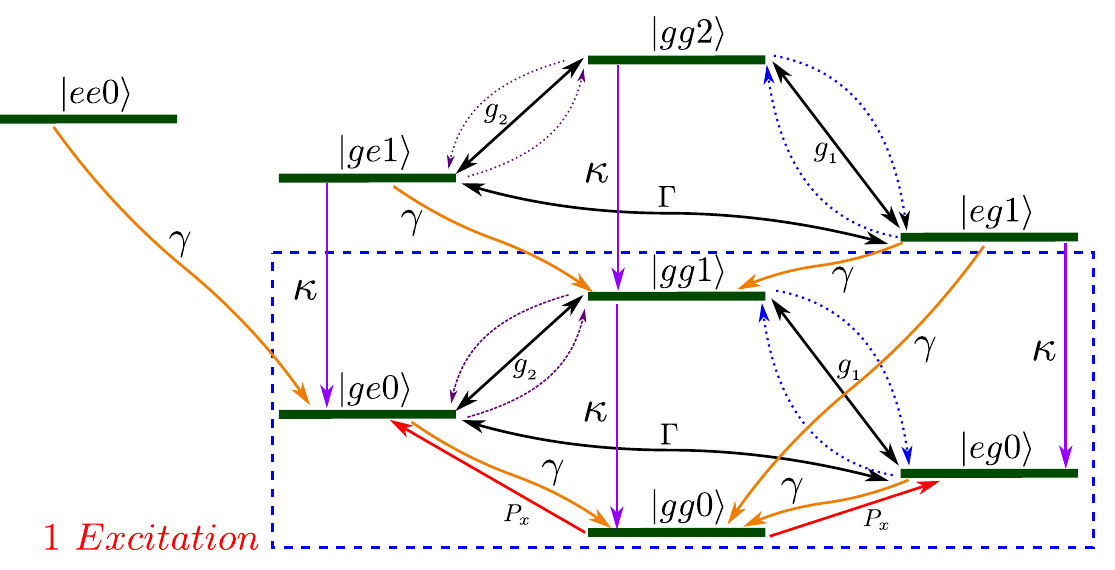}
\caption{Dynamics of a system of two QD's in mutual interaction and interacting with a field mode, in the presence of dissipative and pumping mechanisms. The schematic shows the diagram for two excitations, for Hamiltonian dynamics $\mathcal{H}_{s}$ and non-Hamiltonian dynamics.}
\label{Fig_3}
\end{figure}

Figure \sugb{\ref{Fig_3}} show a schematic representation of the dynamics of the system. We show all the possible transitions between the states of the system up to the second excitation manifold, associated with the Hamiltonian given by Eq.~\sugb{(\ref{Eq_1})}. Considering the first variety of excitation, represented in the box (red dotted lines), it is seen that the Hamiltonian dynamics only admits horizontal transitions between states of the same excitation manifold, and there is no coupling between states belonging to different excitation manifolds. On the other hand, when we take into account the dissipative terms, the non-Hamiltonian dynamics couples states between different excitation manifolds. Notice that, if we consider only the terms $\kappa$ and $\gamma$, the stationary solution will always be the vacuum state $\vert g,g,0\rangle$. Nonetheless, if we introduce the common pumping term, the system can have as a stationary solution populations of excitons or photons different from zero. This is a crucial fact in the determination of the SC as has been observed in different investigations~\sugb{\cite{laucht2010, laucht2009, laussy2008}}

\begin{figure}[t]
\centering
\includegraphics[scale=0.45]{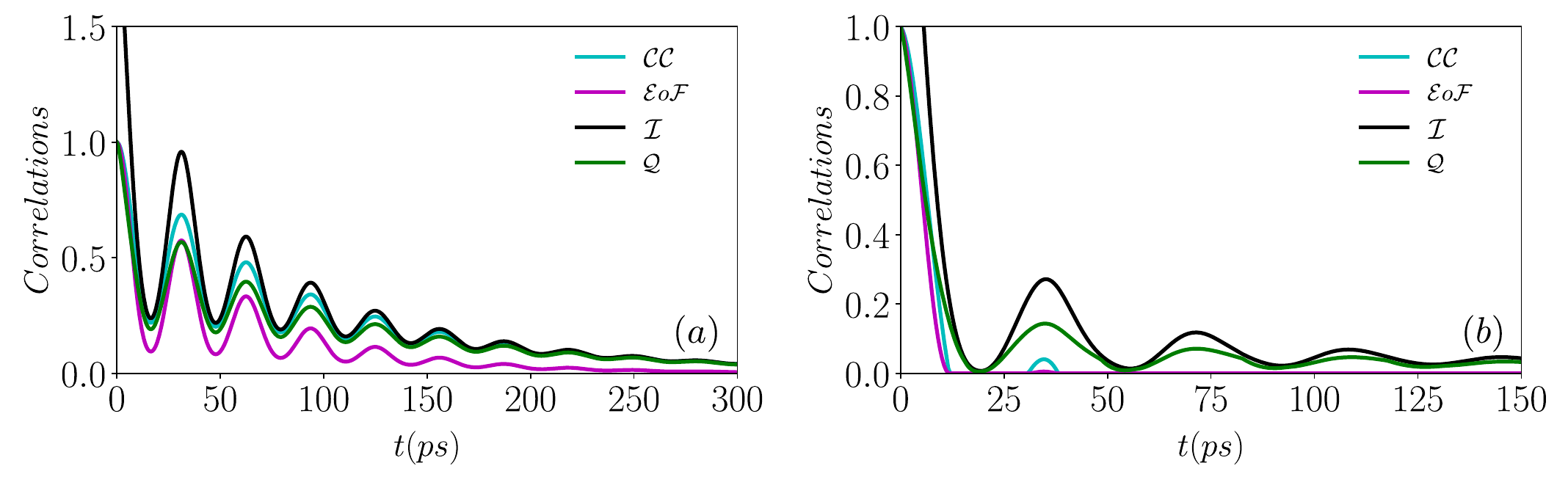}
\caption{Dynamics of the quantum correlations as a function of time in the strong coupling regime for $g/2\pi=10$~GHz, $\gamma/2\pi=0.025$~GHz, $\kappa/2\pi=5$~GHz. (a) considerando $\Gamma/2\pi=15$~GHz and $P_c=P_x=0$; (b) $\Gamma/2\pi=0$ and $P_c/2\pi= 1$~GHz and $P_0/2\pi= 1$~GHz with $\tau_p=20$~ps $(FWHM\sim 47$~ps).}
\label{Fig_4}
\end{figure}

Now we consider the study of quantum correlations as a function of time in the strong coupling regime and considering a short pulse time, where the pumping time $\tau_p$ is shorter ($\tau_p<1/\kappa$) than the useful life of the polariton $1/\kappa$ when the system is pumped by $P_x(t)$ and considering the dissipative effects described by Eq~\sugb{(\ref{Eq_3})}. The Fig.~\sugb{\ref{Fig_4}} shows the dynamics of the quantum correlations for the cases in which the pumping processes are deactivated ($P_c, P_x =0$) and $\Gamma/2\pi= 15$~GHz (Fig.~\sugb{\ref{Fig_4}~(a)}), and for the case in which we activate the pumps ($P_c, P_x \neq 0$) and deactivate the interaction between the QDs, $\Gamma=0$ (Fig.~\sugb{\ref{Fig_4}~(b)}) in which it is observed that the coupling mechanism between the QDs is the interaction with the cavity. For the first case, as shown in Fig.~\sugb{\ref{Fig_4}~(a)}, the calculated quantum correlations show an oscillatory behavior and then gradually disappear over a long time limit. This revival feature is most evident for $\mathcal{E}o\mathcal{F}$. Here, the coherent energy exchange between the QDs due to $\Gamma$ leads to the generation and reactivation of the behavior of quantum correlations that end up disappearing (decay to zero) due to the incessant loss of information in the environment, which results from the spontaneous emission (with rates $\gamma_i$) of the individual QDs.
\medskip

In the case where $\Gamma=0$ and $P_c, P_x\neq 0$ both $\mathcal{I}$ and $\mathcal{Q}$ fade out at a rate relatively slower than $\mathcal{CC}$ and $\mathcal{E}o\mathcal{F}$, but faster than the previous case. In addition, there is an obvious revival for $\mathcal{CC}$ but not for $\mathcal{E}o\mathcal{F}$. Although there is no dipole-dipole interaction in this case, transient quantum correlations can still be established due to the incoherent interactions of correlated spontaneous emissions between the QDs. Despite the fact that concurrency and entanglement disappear, other types of quantum properties are presented even in the system quantized by mutual information and quantum discord that are maintained for a longer time in the system. Therefore, we can see that the Föster interaction acts as a mediation part to induce coherent interactions, so it is expected to be constructive to establish story correlations between QDs or other types of systems.

\begin{figure}[t]
\centering
\includegraphics[scale=0.45]{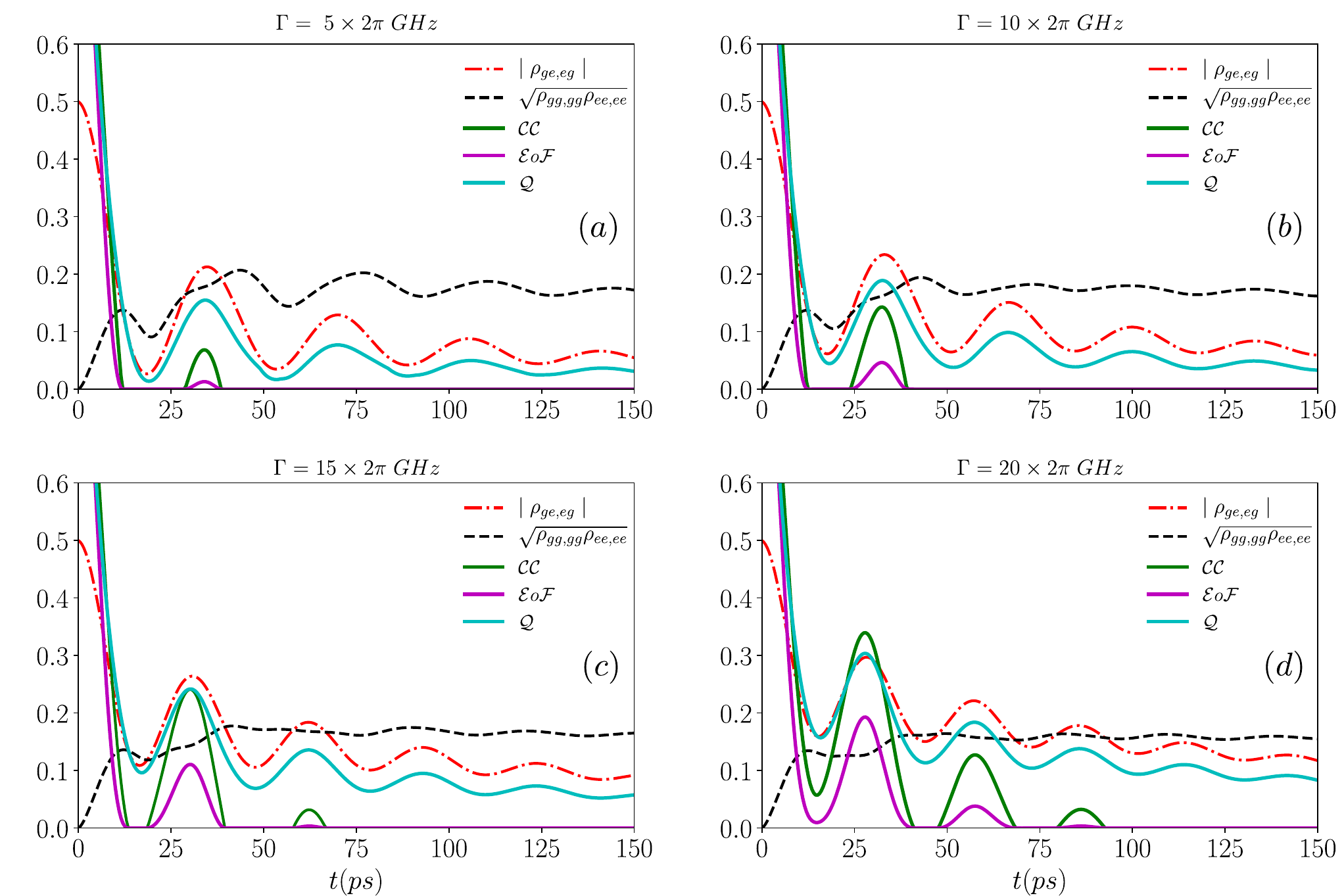}
\caption{Dynamics of the quantum correlations of two atoms as a function of time in the strong coupling regime, under photon pumping $P_C/2\pi=1~GHz$ and a exciton pumping $P_x$ with $\tau_p=20~ps$ (FWHM$\sim 47~ps$), $P_0/2\pi=1GHz$ and different values of the Förster interaction $\Gamma/2\pi=\lbrace{5, 10, 15, 20\rbrace}~GHz$, (a) $\Gamma/2\pi=5~GHz$, (b) $\Gamma/2\pi=10~GHz$,  (c) $\Gamma/2\pi=15~GHz$ and (d) $\Gamma/2\pi=20~GHz$.}
\label{Fig_5}
\end{figure}

Fig.~\sugb{\ref{Fig_4}} shows the behavior of the correlations $\mathcal{CC}$, $\mathcal{E}o\mathcal{F}$, $\mathcal{I}$ and $\mathcal{Q}$, as a function of time for different values of Förster interaction, the values of $\Gamma/2\pi$ are in the range of $5$~GHz to $20$~GHz and also consider $P_c\neq 0$ and a pumping pulse $ P_x$, with $\tau_p=20$~ps and $P_0/2\pi=1$~GHz. The range of parameters for this situation was carefully chosen as large variations in them lead to non-physical results. In fact, if $P_x$ increases carelessly, the basis is not large enough to describe the system, since the average number of photons becomes greater than $1$. Fig.~\sugb{\ref{Fig_4}} shows an interesting issue when observing a discrepancy between $\mathcal{E}o\mathcal{F}$ and $\mathcal{CC}$ as quantifiers of entanglement. Fig.~\sugb{\ref{Fig_4} (a)} shows that concurrency reaches much higher values than $\mathcal{E}o\mathcal{F}$; More surprising is the fact that the concurrency reaches, for almost any moment, values greater than the total correlations, a result that contrasts with the definition of mutual information~$\mathcal{I}$ as a quantity that accounts for all correlations (classical and quantum), including entanglement~\sugb{\cite{hamieh2004, modi2010}}. Therefore, within this framework, concurrency can indicate results that are well above the $\mathcal{E}o\mathcal{F}$, and therefore does not allow direct comparison with other entropic measures such as quantum discord. In addition, the amplitude of the $\mathcal{CC}$ becomes larger and revivals are observed at different time intervals, leading to a non-zero $\mathcal{E}o\mathcal{F}$ amplitude value for various time intervals, since an evident death and a subsequent peak of revival of the correlations $\mathcal{CC}$ and $\mathcal{E}o\mathcal{F}$ can be noted, however, it is appreciated that with the increase from the Förster interaction $\Gamma$ there is a considerable increase in the height of the revival peak of these correlations, as well as an increase in the time in which they remain, which can be verified by comparing with the case in which $\Gamma=0$ as shown in Fig.~\sugb{\ref{Fig_4} (b)}; In addition, for values of $\Gamma>g$ there is a revival of new peaks in the correlations, as well as an increase in their height, which is due to the fact that under this condition the SC regime is guaranteed \sugb{\cite{gomez2019}}.

The reason for this behavior (Fig.~\sugb{\ref{Fig_4}}) is that the concurrency $\mathcal{CC}$ and therefore the entanglement formation $\mathcal{E}o\mathcal{F} $ of our system essentially depends on the difference between the absolute value of the coherence term $\rho_{ge,eg}$ and the square root of the product of the populations $\rho_{gg,gg}$ and $\rho_{ee ,ee}$. Note, for example, that in Fig.~\sugb{{\ref{Fig_4} (a)-(d)}} the $\mathcal{CC}$ and $\mathcal{E}o\mathcal{F}$ disappears precisely for the times that $\vert \rho_{ge,eg}\vert< \sqrt{\rho_{gg,gg}\rho_{ee,ee}}$ and that the difference between the $\mathcal{CC}$ and $\vert\rho_{ge,eg}\vert$ at their minima grows as $\sqrt{\rho_{gg,gg}\rho_{ee,ee}}$ grows. In the dynamics of the density matrix that is considered, it is seen that the coherences decay slower when the value of the Förster interaction increases, on the other hand, $\Gamma$ tends to increase the population of $\rho_{ee,ee}$, while dissipative processes tend to increase the population of $\rho_{gg,gg}$. Even more interesting is the fact that, as dissipation increases, the areas where $CC$ and $\mathcal{E}o\mathcal{F}$ (the so-called sudden entanglement death \sugb{\cite{yu2009, almeida2007}}) widen and the revivals in the concurrency become further apart as the Förster interaction increases. Note that when the Förster interaction decreases, the concurrency maxima are less sharp and eventually disappear, noting that the concurrency is maintained even longer than $\mathcal{E}o\mathcal{F}$.

\begin{figure}[t]
\centering
\includegraphics[scale=0.42]{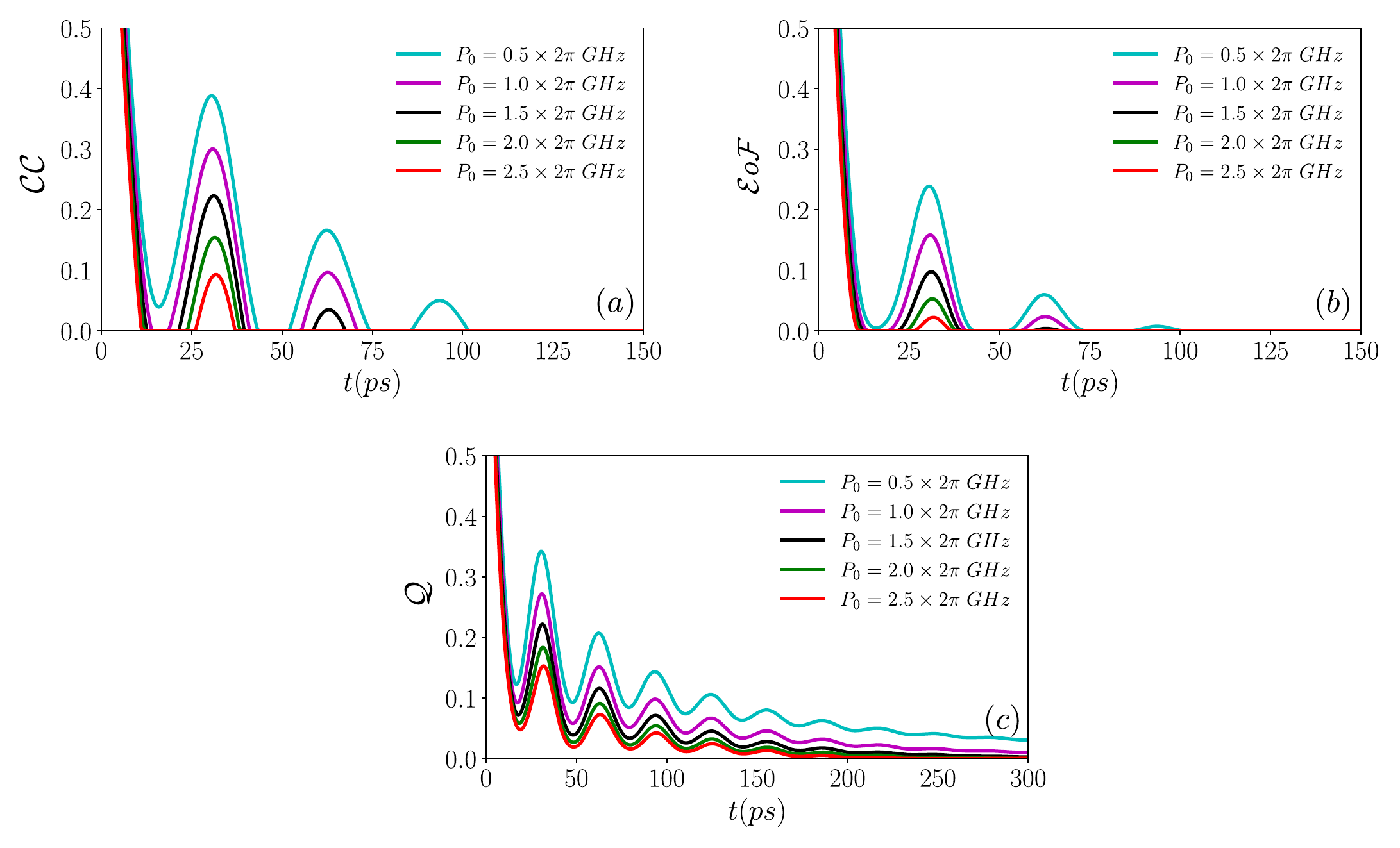}
\caption{Dynamics of the quantum correlations of two atoms as a function of time in the strong coupling regime, under photon pumping $P_C/2\pi=0.5$~GHz, Förster interaction $\Gamma/2\pi=15$~GHz, $ \tau_p=20$~ps and different pump pulse intensities $P_0/2\pi=\lbrace 0.5, 1.0, 1.5, 2.0, 2.5\rbrace$~GHz for, (a) Concurrence $\mathcal{CC}$, (b) formation entanglement $\mathcal{E}o\mathcal{F}$, (c) Quantum Discord $\mathcal{Q}$.}
\label{Fig_6}
\end{figure}

On the other hand, in order to observe the pumping effects of the excitonic pulse $P_x$ on the system, Fig.~\sugb{\ref{Eq_6}} and Fig.~\sugb{\ref{Eq_7}} are considered, showing the dynamics of the quantum correlations $\mathcal{CC}$, $\mathcal{E}o\mathcal{F}$ and $\mathcal{Q}$, for $\Gamma/2\pi=15$~GHz and for different intensities of the pump pulse $P_0$ (Fig.~\sugb{\ref{Eq_6}}), and different time scales of the pump $\tau_p$ (Fig.~\sugb{\ref{Eq_7}}). A damped oscillatory behavior is observed in the correlations, due to the presence of loss processes at the rates $\kappa$, and $\gamma$, a similar result is reported by \sugb{\cite{zhang2012}}. In Fig.~\sugb{\ref{Eq_6}~(a)-(b)} three maximum entanglement peaks are observed for the case of $P_0/2\pi=0.5$~GHz, one around $30$~ps, another at $62$~ps and another at $94$~ps, indicating that the greatest entanglement of the system occurs at these moments of time. However, the quantum discord shows values different from zero at all times that disappear for very long times and that are reduced as the value of the intensity of the pumping pulse increases. On the other hand, in all cases, the first maximum peak of $\mathcal{CC}$ and $\mathcal{E}o\mathcal{F}$ is greater than the second and is almost completely maintained at the same time. for increments of $P_0$. However, it is observed that the second and third peaks disappear as the intensity of the pulse is increased.
\medskip

\begin{figure}[t]
\centering
\includegraphics[scale=0.42]{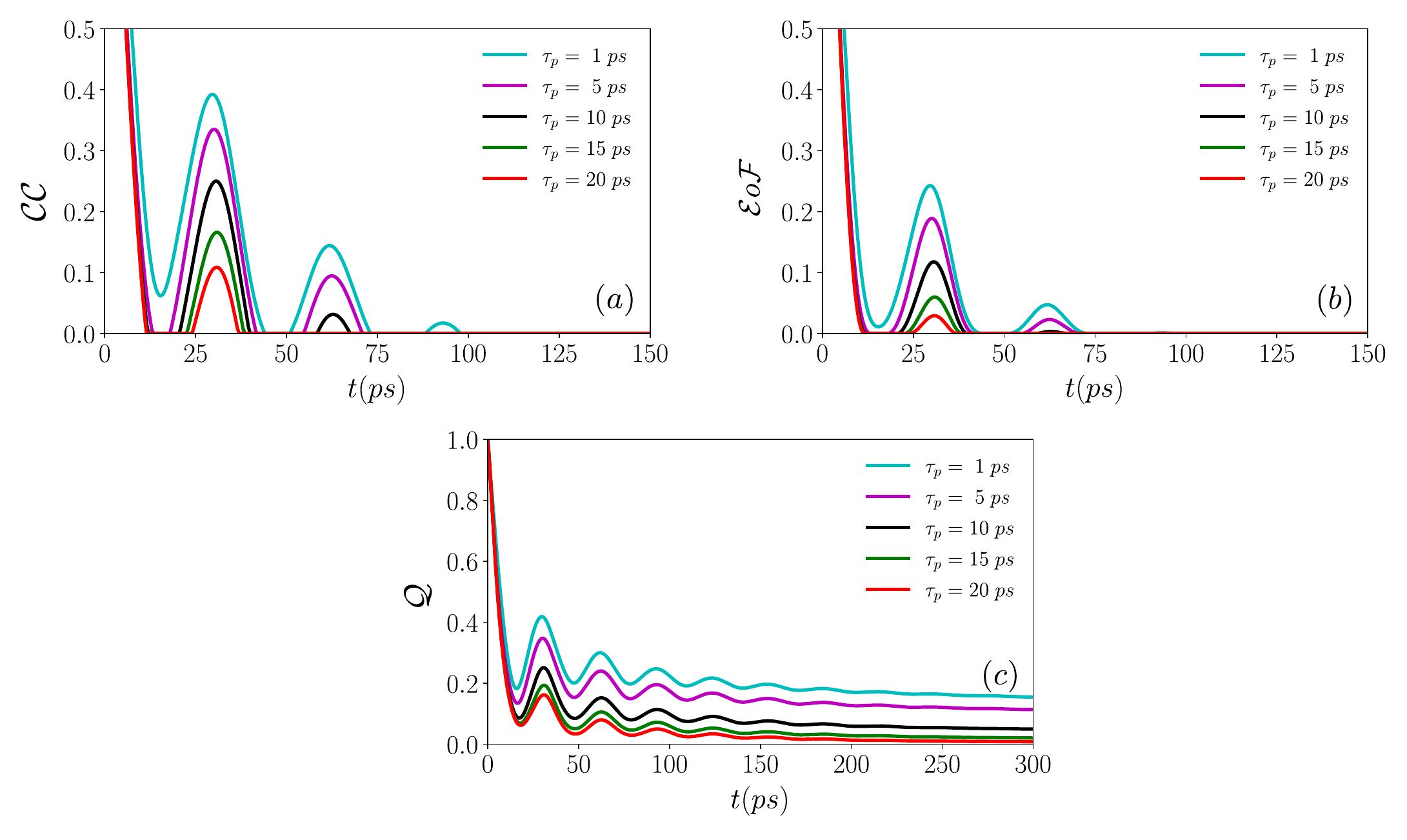}
\caption{Dynamics of the quantum correlations of two atoms as a function of time in the strong coupling regime, under photon pumping $P_C/2\pi=1.0$~GHz and an exciton pumping $P_x$ with $P_0/2\pi=1$~GHz, $\Gamma/2\pi=15$~GHz and different pump pulse times $\tau_p=\lbrace{1, 5, 10, 15, 20 \rbrace}$~ps for, (a) Concurrence~$\mathcal{CC}$, (b) Quantum Entanglement Formation~$\mathcal{E}o\mathcal{F}$, and (c) Quantum Discord $\mathcal{Q}$.}
\label{Fig_7}
\end{figure}

When considering the variation of $\tau_p$, as shown in Fig.~\sugb{\ref{Eq_7}}, the damped oscillatory behavior of Fig.~\sugb{\ref{Eq_6}} is still observed characterized by the two or three peaks depending on the value $\tau_p$, but in this case the damping is faster only for the quantum discord, which for values of $\tau_p$ between $1$~ps and $10$~ps reaches Asymptotic values different from zero, while for $\mathcal{CC}$ and $\mathcal{E}o\mathcal{F}$ no significant changes are observed. Note that with $\tau_p=1$~ps the oscillations of $\mathcal{CC}$ and $\mathcal{E}o\mathcal{F}$ disappear around $75$~ps and as $\tau_p$ increases, the time in which they disappear decreases considerably so that when $\tau_p=20$~ps disappear at around $30$~ps. Also, as $\tau_p$ increases, the valley between the peaks increases considerably, causing the second and third peaks to decrease in height with with respect to the first until they completely disappeared.

These effects can be explained by looking at the time duration of the first two collapses of $\mathcal{CC}$ and $\mathcal{E}o\mathcal{F}$ ($\delta t1$ is the time interval of the first collapse and $\delta t2$ the time interval of the second). It is clear that as non-Hamiltonian effects become more important, the length of both intervals gets longer and that for any $P_0$ and $\tau_p$ the second interval $\delta t2$ is wider than the first $\delta t1$. This behavior can be understood as follows: as explained above, the degree of entanglement between subsystems is the difference $\vert\rho_{ge,eg}\vert-\sqrt{\rho_{gg,gg}\rho_ {ee,ee}}$ and the dynamic behavior of the coherence term is decreasing with the rate of decrease proportional to the sum of $\kappa$ and $P_x$. To have entanglement the absolute value of $|\rho_{ge,eg}|$ must be greater than $\sqrt{\rho_{gg,gg}\rho_{ee,ee}}$. Now notice that as $P_0$ increases, the oscillations become more damp. Due to damping, the amplitude of the oscillations is smaller and the absolute value of the coherence ($\rho_{ge,eg}$) has to be at a time $t$ closer to a maximum to be greater than $\sqrt{ \rho_{gg,gg}\rho_{ee,ee}}$, that is, the finite time where there is no entanglement approximates the time interval between two successive maximums of $\vert\rho_{ge,eg}\vert $ as $P_0$ is increased. To understand the fact that the first sudden death interval $\delta t1$ is smaller than the second $\delta t2$, simply note that for times in the second interval the amplitude of $\vert\rho_{ge, eg}\vert$ will be smaller than in the first (Fig~{\sugb{\ref{Eq_6}} and \sugb{\ref{Eq_7}}), so the times at which entanglement will be nonzero will be closer to the maximums of $\vert\rho_{ge ,eg}\vert$ resulting in a wider interval.

This behavior suggests that by manipulating experimentally accessible parameters, such as the Förster interaction, laser pump intensity, and pump pulse time, control of the quantum entanglement and discord of the system can in principle be obtained.

\section{Conclusions}\label{Conclusions}

In this work, we have investigated concurrence variation $\mathcal{CC}$, entanglement formation $\mathcal{E}o\mathcal{F}$, and discord $\mathcal{Q}$ in the matrix of two coupled quantum dots that interact with a cavity mode and are pumped with a laser pulse. Each quantum dot has an exciton that can be modeled by an electric dipole. We have shown that the number of quantum correlations increases with increasing Förster interaction and decreases each time we activate exciton pumping either by increasing the intensity or the pulse time.

Furthermore, a discrepancy between $\mathcal{E}o\mathcal{F}$ and $\mathcal{CC}$ as entanglement quantifiers is shown, noting that concurrence reaches much higher values (and decays much more slowly) than $\mathcal{E}o\mathcal{F}$; whereby concurrence may indicate results that are well above the $\mathcal{E}o\mathcal{F}$ and thus does not allow direct comparison with other entropic measures such as quantum discord. This shows that care must be taken when confronting or using entanglement-specific quantifiers to describe a physical process, especially since the interpretation of the $\mathcal{E}o\mathcal{F}$ as the cost of creating an entangled state leads to an upper bound on the actual degree of entanglement in a quantum system~\sugb{\cite{cornelio2011}}. It has also been shown that the system maintains quantum correlations even when the entanglement of the system has disappeared, given by the quantum discord. Quantum discord is the dominant correlation that prevails throughout the dissipative dynamics and that this is enhanced by the presence of the Förster interaction, but not by exciton laser pumping.

\section*{Acknowledgment}

A.A. Portacio, acknowledges financial support from Universidad de los Llanos under Project "Cálculo de la generación de segundo armónico en nanoestructuras usando una ecuación maestra en la forma de Lindblad" code C07-F02-004 2022. And "Cálculo de la respuesta óptica no lineal de una molécula de dos puntos cuánticos inmersos en una nanocavidad semiconductora (QDM-Cavity system)" code C09-F02-011 2022.
D.A. Rasero, thanks Universidad Surcolombiana for the support during the completion of this work.
D. Madrid, thanks the Universidad de Sucre for the support during the completion of this work.

%\appendix
%\section{Zzzz}\label{App1}

\printcredits

%% Loading bibliography style file
\bibliographystyle{model1-num-names}

% Loading bibliography database
\bibliography{cas-refs}

%\newpage
%\vspace{1.0truecm}
%\vskip3pt

%\bio{}
%Author biography without author photo.
%Author biography. Author biography. Author biography.
%\endbio

%\vspace{1.0truecm}
%\bio{figs/pic1}
%Author biography without author photo.
%Author biography. Author biography. Author biography.
%\endbio

%\bio{figs/pic1}
%Author biography without author photo.
%Author biography. Author biography. Author biography.
%\endbio

%\bio{figs/pic1}
%Author biography without author photo.
%Author biography. Author biography. Author biography.
%\endbio

%\bio{figs/pic1}
%Author biography without author photo.
%Author biography. Author biography. Author biography.
%\endbio

\end{document}